\documentclass[preprint,preprintnumbers,showpacs,aps,amssymb]{revtex4-1}

\usepackage{graphicx}
\usepackage{bm}
\usepackage{amsmath}

\def\MBH{M_{\rm BH}}

\def\td{{\tilde\delta}}
\def\tE{{\tilde E}}
\def\tL{{\tilde\Lambda}}
\def\tR{{\tilde R}}
\def\tw{{\tilde\omega}}

\def\calO{{\cal O}}

\def\wmax{\omega_{\rm max}}
\def\r0min{r_{0,{\rm min}}}

\def\nn{\nonumber}

\begin{document}
\title{Superradiance by mini black holes with mirror}
\author{Jong-Phil Lee}
\email{jplee@kias.re.kr}
\affiliation{Department of Physics and IPAP, Yonsei University, Seoul 120-749, Korea}
\affiliation{Division of Quantum Phases $\&$ Devices, School of Physics, Konkuk University, Seoul 143-701, Korea}

\begin{abstract}
The superradiant scattering of massive scalar particles by a rotating mini black hole is investigated. 
Imposing the mirror boundary condition, the system becomes the so called black-hole bomb 
where the rotation energy of the black hole is transferred to the scattered particle exponentially with time.
Bulk emissions as well as brane emissions are considered altogether.
It is found that the largest effects are expected for the brane emission of lower angular modes
with lighter mass and larger angular momentum of the black hole.
Possibilities of the forming the black-hole bomb at the LHC are discussed.
\end{abstract}

\maketitle
\section{Introduction}
The standard model (SM) of the elementary particles is now undergoing the greatest challenges ever by the successful
running of the Large Hadron Collider (LHC) at CERN.
The last missing piece of the SM, the Higgs boson, would be discovered or ruled out sooner or later and new physics signals
beyond the SM are expected to appear for the first time.
Among the most striking events that the LHC would reveal is the formation of mini black holes.
If there are extra dimensions \cite{XD} the fundamental energy scale for strong gravitation $M_D$ in $D=(4+n)$-spacetime
dimensions can be much smaller than the Planck mass.
In the case of $M_D\sim 1$ TeV, the LHC experiments can probe the strong gravity regime and possibly produce black holes
\cite{BH}.
Typical size of the Schwarzschild radius of these black holes is $\sim\calO(10^{-4})$ fm and they are called as mini black holes.
Mini black holes also evaporate through Hawking radiation \cite{Hawking} just as ordinary black holes.
The lifetime of a mini black hole is typically about $\sim 10^{-26}~\sec$ so the black hole decays instantaneously.
Once the black hole is produced at the LHC one can detect its creation by observing Hawking radiation.
It is also possible that rotating or charged black holes would be produced.
For these reasons there have been many studies on the emission of particles by mini black holes, both static and rotating ones
in extra dimensions \cite{Frolov,Ida,Harris,Pappas}.
\par
When particles are scattered off by rotating (or Kerr) black holes a very interesting phenomenon known as superradiance occurs.
If a field of the form $e^{-i\omega t}e^{im\phi}$ is incident on a rotating black hole of angular velocity $\Omega$,
the field is amplified if $\omega<m\Omega$.
In this case the rotation energy is transferred to the scattered field.
For any rotating objects the superradiance can occur.
If the rotating black hole is surrounded by a mirror which reflects the scattered field back to the black hole, 
the field bounces back and forth between the black hole and the mirror.
During the process the field amplifies itself and the extracted energy from the rotating black hole grows exponentially.
This is so called the Press-Teukolsky black-hole bomb \cite{Press}.
The origin of the black-hole bomb is the small imaginary part of $\omega$.
In the case of superradiance where $\omega<m\Omega$, the imaginary part is positive so the field contains the term of
exponential increasing with time, $\sim e^{({\rm Im}~\omega) t}$.
The mirror which reflects the field needs not be a real one.
For example, a mass term of the field in 4-dimensional spacetime can play a role of the mirror.
In this case the massive field forms a bound state with the black hole, and the superradiance results in the instability
of the black hole.
There have been many works on the black-hole bomb where a massive particle is bound to a celestial black hole 
under superradiance \cite{Cardoso, Dolan, Hod,Rosa}.
For the higher dimensional Kerr black holes, it was shown that they do not exhibit such instabilities for massive fields 
because there are no bound geodesics in the spacetime around the black hole \cite{Yoshida}.
\par
In a recent work \cite{jplee} I have investigated the possibility of black-hole bombs at the LHC.
The work is a higher dimensional version of \cite{Cardoso}.
In \cite{jplee} the superradiance of a massive scalar field by a rotating black hole in higher dimensions with the mirror boundary
condition is analyzed.
As shown in \cite{Yoshida} the mass term would not be enough for the role of mirror and probably electromagnetic fields might
work out for charged particles, but to focus on the possibility of forming a black-hole bomb, the mirror boundary condition was simply imposed 
without considering any technical concerns about it.
It was assumed that the mirror boundary condition is simply given by whatever reflects the scattered particle back to the black hole.
Current work is an extension of \cite{jplee}.
In this work we analyzed the bulk emission as well as the brane emission.
Intuitively the bulk emission modes are expected to be suppressed compared with the brane emission, 
and we will see that this is really the case by a factor of $\lesssim 10^{-2}$.
We also investigate the different angular modes for different masses.
\par
In the next Section the superradiance of massive scalar particles by rotating black holes in higher dimensions 
with the mirror boundary condition is described.
Both of the bulk emission and the brane emission are considered.
In Sec.\ III the results are summarized and discussions are given.
The conclusions appear in Sec.\ IV.
\section{Rotating black holes in extra dimensions and scalar emissions}
In this section, we describe the rotating black holes in higher $4+n$ dimensions  just as in the way of \cite{Pappas}.
The procedure is much the same as that of \cite{Pappas,jplee}.
We assume that the angular momentum lies parallel to our brane.
A proper background geometry is the Myers-Perry solution \cite{MP}
\begin{eqnarray}
 ds^2
&=&
-\left(1-\frac{\mu}{\Sigma~r^{n-1}}\right)dt^2 -\frac{2a\mu\sin^2\theta}{\Sigma~r^{n-1}}dtd\phi
+\frac{\Sigma}{\Delta}dr^2 \nn\\
&&
+\Sigma d\theta^2+\left(r^2+a^2+\frac{a^2\mu\sin^2\theta}{\Sigma~r^{n-1}}\right)\sin^2\theta d\phi^2\nn\\
&&
+r^2\cos^2\theta d\Omega_n^2~,
\label{MP}
\end{eqnarray}
where
\begin{equation}
 \Delta=r^2+a^2-\frac{\mu}{r^{n-1}}~,~~~\Sigma=r^2+a^2\cos^2\theta
\end{equation}
and $d\Omega_n^2$ is a unit $n$-sphere.
The mass parameter $\mu$ is related to the black hole mass $\MBH$ and the new fundamental scale $M_D$ in $D=4+n$ dimensions via
\begin{equation}
 \mu=\frac{\Gamma(\frac{n+3}{2})}{(n+2)\pi^{\frac{n+3}{2}}} (2\pi)^n \frac{\MBH}{M_D^{n+2}}~,
\end{equation}
and $a$ is proportional to the black hole angular momentum $J$ as
\begin{equation}
 J=\frac{2}{n+2}\MBH a~.
\end{equation}
The radius of the black hole horizon $r_h$ is determined by
\begin{equation}
 r_h^{n+1}=\frac{\mu}{(1+a_*^2)}~,
\end{equation}
which makes $\Delta=0$, and $a_*=a/r_h$.
\par
A massive scalar field $\Phi$ of mass $m_0$ under the gravitational background (\ref{MP}) satisfies 
the Klein-Gordon equation in curved spacetime
\begin{equation}
 \frac{1}{\sqrt{-G}}\partial_A\left(\sqrt{-G}G^{AB}\partial_B\Phi\right)-m_0^2\Phi=0~,
\label{KG}
\end{equation}
where $G_{AB}$ is the $D$-dimensional metric tensor of (\ref{MP}) and $G$ is its determinant.
The equation can be solved by the separation of variables for $\Phi$ as
\begin{equation}
 \Phi=e^{-i\omega t}e^{im\phi}R(r)S(\theta)Y_{\ell n}(\theta_1,\cdots,\theta_{n-1},\phi)~,
\end{equation}
where $Y_{\ell n}(\theta_1,\cdots,\theta_{n-1},\phi)$ are the hyperspherical harmonics on the $n$-sphere.
The radial function $R(r)$ satisfies 
\begin{equation}
 \frac{1}{r^n}\partial_r (r^n\Delta\partial_r R)+\left(
 \frac{K^2}{\Delta}-\frac{\ell(\ell+n-1)a^2}{r^2}-\tL_{j\ell m}-m_0^2r^2\right)R=0~,
\end{equation}
while the angular function $S(\theta)$ satisfies
\begin{equation}
 \frac{1}{\sin\theta\cos^n\theta}\partial_\theta (\sin\theta\cos^n\theta\partial_\theta S) +\left(
 \tw^2a^2\cos^2\theta-\frac{m^2}{\sin^2\theta}-\frac{\ell(\ell+n-1)}{\cos^2\theta}+\tE_{j\ell m}\right)S=0~,
\end{equation}
where
\begin{eqnarray}
\tw&=&\sqrt{\omega^2-m_0^2}~,\\
 K&=&(r^2+a^2)\omega-am~,\\
 \tL_{j\ell m}&=&\tE_{j\ell m}+a^2\omega^2-2am\omega~,
\end{eqnarray}
and $\tE_{j\ell m}$ is expanded around small $a\tw$ as
\begin{eqnarray}
\tE_{j\ell m}&=&j(j+n+1) \nn\\
&&-(a\tw)^2\frac{-1+2\ell(\ell-1)+2j(j+1)-2m^2+2n(n+\ell)+n^2}{(2j+n-1)(2j+n+3)}+\calO((a\tw)^4)~.
\label{tE}
\end{eqnarray}
\par
To investigate the superradiance regime, it is enough to concentrate on the radial function $R(r)$
since the absorption and reflection probabilities of scattering are determined by $R(r)$.
To solve the radial equation, first consider the near-horizon region where $r\simeq r_h$. 
After changing the variable as $r\to f(r)=\Delta(r)/(r^2+a^2)$,  the radial equation becomes 
\begin{equation}
 f(1-f)\frac{d^2R}{df^2}+(1-D_*f)\frac{dR}{df}+\left[
 \frac{K_*^2}{A_*^2f(1-f)}-\frac{C_*}{A_*^2(1-f)}\right]R=0~,
\label{Rf}
\end{equation}
where
\begin{eqnarray}
 A_*&=&(n+1)+(n-1)a_*^2~,\\
K_*&=&(1+a_*^2)\omega r_h-a_*m~,\\
C_*&=&\big[\ell(\ell+n-1)a_*^2+\tL_{j\ell m}+m_0^2r_h^2\big](1+a_*^2)~,\\
D_*&=&1-\frac{4a_*^2}{A_*^2}~.
\end{eqnarray}
The solution for $R=R(f)$ in Eq.\ (\ref{Rf}) is proportional to the hypergeometric function $F(a,b,c;f)$ as
\begin{equation}
 R_{\rm NH}(f)=A_-f^\alpha(1-f)^\beta F(a,b,c;f)~,
\label{RNH}
\end{equation}
where 
\begin{eqnarray}
 \alpha&=&-i\frac{K_*}{A_*}~,\\
 \beta&=&\frac{1}{2}\left[(2-D_*)-\sqrt{(2-D_*)^2-\frac{4(K_*^2-C_*)}{A_*^2}}~\right]~,
\end{eqnarray}
with
\begin{equation}
 a=\alpha+\beta+D_*-1~,~~~b=\alpha+\beta~,~~~c=1+2\alpha~,
\end{equation}
and $A_-$ is the integration constant.
\par
In the far-field region where $r\gg r_h$, the radial equation becomes
\begin{equation}
 \frac{d^2\tR}{dz^2}+\frac{1}{z}\frac{d\tR}{dz}+\left[
 1-\frac{\tE_{j\ell m}+a^2\tw^2+\left(\frac{n+1}{2}\right)^2}{z^2}\right]\tR=0~,
\end{equation}
where $\tR(r)\equiv r^{(n+1)/2} R(r)$ and $z\equiv \tw r$, and the solution is the Bessel function,
\begin{equation}
 R_{\rm FF}(r)=\frac{B_1}{r^{(n+1)/2}}J_\nu(\tw r)+\frac{B_2}{r^{(n+1)/2}}Y_\nu(\tw r)~.
\label{RFF}
\end{equation}
Here $J_\nu$ and $Y_\nu$ are the first and second kind of the Bessel functions, 
where 
\begin{equation}
\nu=\sqrt{\tE_{j\ell m}+a^2\tw^2+(n+1)^2/4}~.
\end{equation}
Matching the two solutions $R_{\rm NH}$ and $R_{\rm FF}$ gives the ratio of $B_2$ to $B_1$ as \cite{Pappas}
\begin{eqnarray}
 \frac{B_2}{B_1}
&=&-\pi\left[\frac{\tw r_h(1+a_*^2)^{1/(n+1)}}{2}\right]^{2j+n+1}
 \frac{1}{\nu\Gamma^2(\nu)}\nn\\
&&\times
 \frac{\Gamma(2\beta+D_*-2)\Gamma(2+\alpha-\beta-D_*)\Gamma(1+\alpha-\beta)}
        {\Gamma(\alpha+\beta+D_*-1)\Gamma(\alpha+\beta)\Gamma(2-2\beta-D_*)}~.
\label{match}
\end{eqnarray}
\par
The results up to now hold for any scalar particles embedded in the geometry of (\ref{MP}) through Eq.\ (\ref{KG}).
Our concern is the system surrounded by a mirror which reflects the scalar particle back to the black hole.
If the mirror is located at $r=r_0$, then the mirror boundary condition is $R_{\rm FF}(r_0)=0$, which means \cite{Cardoso}
\begin{equation}
 \frac{B_1}{r_0^{(n+1)/2}}J_\nu(\tw r_0)+\frac{B_2}{r_0^{(n+1)/2}}Y_\nu(\tw r_0)=0~.
\label{BC}
\end{equation}
Combining Eqs.\ (\ref{match}) and (\ref{BC}) gives
\begin{eqnarray}
 \frac{J_\nu(\tw r_0)}{Y_\nu(\tw r_0)}
&=&\pi\left[\frac{\tw r_h(1+a_*^2)^{1/(n+1)}}{2}\right]^{2j+n+1}
 \frac{1}{\nu\Gamma^2(\nu)}\nn\\
&&\times
 \frac{\Gamma(2\beta+D_*-2)\Gamma(2+\alpha-\beta-D_*)\Gamma(1+\alpha-\beta)}
        {\Gamma(\alpha+\beta+D_*-1)\Gamma(\alpha+\beta)\Gamma(2-2\beta-D_*)}~.
\label{BC2}
\end{eqnarray}
In the limit of $\tw_*\equiv\tw r_h\ll 1$ the right-hand-side of Eq.\ (\ref{BC2}) is 
\begin{equation}
 \frac{J_\nu(\tw r_0)}{Y_\nu(\tw r_0)}\sim(\tw_*)^{2j+n+1}\simeq 0~,
\end{equation}
so that $J_\nu(\tw r_0)\simeq 0$ in this limit.
To a good approximation, the value of $\tw r_0$ is equal to the zeros of the Bessel function $J_\nu(x)$,
\begin{equation}
 \tw r_0\simeq x_{\nu,k}~,
\label{Besselzero}
\end{equation}
where $J_\nu(x_{\nu,k})=0$.
Assume that ${\rm Re}(\tw)\gg {\rm Im}(\tw)$, the solution of Eq.\ (\ref{BC2}) now can be written as 
\begin{equation}
 \tw=\frac{x_{\nu,k}}{r_0}+\frac{i\td}{r_0}~,~~~|\td|\ll 1~,
\end{equation}
where $\td/r_0\equiv\delta$ is introduced as a small imaginary part of $\tw$.
For small $\td$,  $J_\nu(\tw r_0)=J_\nu(x_{\nu,k}+i\td)\simeq i\td J'_\nu(x_{\nu,k})$, 
and one can extract the expression for $i\td$:
\begin{eqnarray}
i\td
 &=&\frac{Y_\nu({x_{\nu,k}})}{J'_\nu(x_{\nu,k})}
\pi\left[\frac{\tw_*(1+a_*^2)^{1/(n+1)}}{2}\right]^{2j+n+1}
 \frac{1}{\nu\Gamma^2(\nu)}\nn\\
&&\times
 \frac{\Gamma(2\beta+D_*-2)\Gamma(2+\alpha-\beta-D_*)\Gamma(1+\alpha-\beta)}
        {\Gamma(\alpha+\beta+D_*-1)\Gamma(\alpha+\beta)\Gamma(2-2\beta-D_*)}~.
\label{td}
\end{eqnarray}
On the right-hand-side of the above equation the imaginary part comes from the terms containing $\alpha=-iK_*/A_*$.
The right-hand-side also has a nonzero real part.
It can be considered as a correction to Eq.\  (\ref{Besselzero}), and for the brane emission the size of it is typical around 1\%.
For the bulk emission we found that it is much more suppressed because of the factor 
$(\tw_*)^{2j+n+1}$, so it will be neglected.
Using the properties of the Gamma functions, one arrives at the final result
\begin{eqnarray}
 \td
 &=&\frac{Y_\nu({x_{\nu,k}})}{J'_\nu(x_{\nu,k})}
\left[\frac{\tw_*(1+a_*^2)^{1/(n+1)}}{2}\right]^{2j+n+1}
\frac{1}{2}\sinh\pi\left(\frac{2K_*}{A_*}\right)\nn\\
 &&\times
\frac{\left[\beta^2+\frac{K_*^2}{A_*^2}\right]\left|\Gamma\left(-\beta+i\frac{K_*}{A_*}\right)\right|^2
\left[(1-\beta-D_*)^2+\frac{K_*^2}{A_*^2}\right]\left|\Gamma\left(1-\beta-D_*+i\frac{K_*}{A_*}\right)\right|^2}
{\nu\Gamma^2(\nu)(2-2\beta-D_*)\Gamma^2(2-2\beta-D_*)}~.\nn\\
\end{eqnarray}
The prefactor of $Y_\nu(x_{\nu,k})/J'_\nu(x_{\nu,k})$ is negative for our relevant range of $\omega$,
so the sign of $\td$ is opposite to that of $K_*=(1+a_*^2)\omega r_h-a_*m$.
(Especially, for small $K_*$, $\td\sim -K_*$.)
More specifically, $\td>0$ for $(1+a_*^2)\omega r_h<a_*m$, which is the superradiant condition.
For the brane emission, one can follow the procedure of \cite{jplee} and use its results.
\par
It is not difficult to see that the energy emission rate of the black hole $dE^2/dtd\omega$ is proportional to $\delta$ \cite{Pappas}.
Thus a larger value of $\delta$ means the stronger instability of the black hole via superradiance.
\section{Results and discussions}
In our analysis we fix $M_D=1$ TeV and $\MBH=5$ TeV.
The relevant range of $\omega$ for the superradiance is
\begin{equation}
 m_0<\omega<\frac{a_*m}{(1+a_*^2)r_h}\equiv\wmax~.
\label{wmax}
\end{equation}
Since $\tw r_0=r_0\sqrt{\omega^2-m_0^2}\simeq x_{\nu,k}$ from the mirror boundary condition (\ref{Besselzero}), one has
\begin{equation}
 r_0>\frac{x_{\nu,k}}{\sqrt{\wmax^2-m_0^2}}\equiv \r0min~.
\label{r0min}
\end{equation}
We only consider the case of $k=1$ for simplicity.
The size of the horizon radius $r_h$ is  $r_h\simeq (2-6)\times 10^{-4}~{\rm fm}$ for $n=2,\cdots,7$ and $a_*=0.5,~1.0,~1.5$.
As for $\r0min$, a dimensionless parameter $\r0min/r_h$ is almost $n$-independent for the brane emission with sufficiently small $m_0$.
But in the case of bulk emission, the parameter $\nu$ itself is $n$-dependent.
\par
Figure 1 shows $\delta$ as a function of $\omega r_h$ in the bulk emission for $m_0=m_\pi$, $m_0=120$ GeV and
for the modes $(j\ell m)=(101), (202)$,
and the details of information are summarized in Tables \ref{bulk101pi}-\ref{bulk202120}.
In these Tables, $\omega_c$ is the value of $\omega$ which makes $\delta$ maximum, 
so $\delta_{\rm max}\equiv\delta(\omega_c)$, and $t_c=1/\delta_{\rm max}$ is the $e$-folding time.
Note that $\delta_{\rm max}$ gets larger for smaller $n$ and for larger $a_*$ (but when $a_*=1.5$ one should be cautious; see below.), 
and for lighter mass $m_0$.
The suppression of $\delta_{\rm max}$ is severe for large $n$ because $\td\sim(\tw_*)^{2j+n+1}$.
Also $\delta_{\rm max}^{(j\ell m=101)}<\delta_{\rm max}^{(j\ell m=202)}$ generally, 
with the only exception of $\delta_{\rm max}^{(101)}(n=2, a_*=0.5)>\delta_{\rm max}^{(202)}(n=2, a_*=0.5)$.
But it must be noted that for $(j\ell m)=(202)$ mode with $a_*=1.5$, the value of $a_*\tw_*^{\rm max}=1.38~(1.37)$
for $m_0=m_\pi~(120~{\rm GeV})$. 
($\tw_*$ itself is $\tw_*=0.923~(0.989)$ in this case.)
For the region $a_*\tw_*>1$ the series of Eq.\ (\ref{tE}) is not a good expansion so we cannot rely on the validity 
of the whole analysis.
For $a_*=1.5$, the analysis is reliable when $\omega r_h<\sqrt{m_0^2 r_h^2+1/a_*^2}$.
In this case the maximum value of $\delta$ is obtained for $\omega r_h=\sqrt{m_0^2 r_h^2+1/a_*^2}$.
In Tables \ref{bulk202pi} and \ref{bulk202120} all the parameters for $a_*=1.5$ are calculated for this region.
For higher modes of $(j\ell m)$, the value of $a_*\tw_*^{\rm max}$ exceeds unity even for $a_*=1$, 
so we do not consider the higher modes.
\par
As a comparison, we provide the results for the emission ``on the brane'' in Fig.\ \ref{brane} and 
Tables \ref{brane11120}-\ref{brane22120}. 
(For the details of the brane emission, see \cite{jplee}).
In general, $\delta_{\rm max}$ on the brane is much larger (about $\calO(10^2)$) than that in the bulk, 
since in the case of brane emission $\td\sim(\tw_*)^{2j+1}$.
As in the bulk emission, $\delta_{\rm max}$ gets larger for smaller $n$ and for larger $a_*$, and for lighter mass $m_0$.
One exception is that $\delta_{\rm max}^{(22)}(n=2, a_*=0.5)<\delta_{\rm max}^{(22)}(n=3, a_*=0.5)$ for $m_0=m_\pi$.
One different feature of brane emission is that $\delta_{\rm max}^{(jm=11)}<\delta_{\rm max}^{(jm=22)}$ mode
for $m_0=120$ GeV (one exception is $\delta_{\rm max}^{(11)}(n=2, a_*=0.5)>\delta_{\rm max}^{(22)}(n=2, a_*=0.5)$)
while $\delta_{\rm max}^{(jm=11)}>\delta_{\rm max}^{(jm=22)}$ for $m_0=m_\pi$.
In the brane emission, one encounters the same problem when $a_*=1.5$ for $(jm)=(22)$, as in the bulk emission.
The value of $\omega$ should be restricted as $\omega r_h<\sqrt{m_0^2 r_h^2+1/a_*^2}$,
and the maximum value of $\delta$ is given when $\omega r_h=\sqrt{m_0^2 r_h^2+1/a_*^2}$.
See Tables \ref{brane22pi} and \ref{brane22120}.
Higher $(jm)$ modes are not considered as before.
In the current analysis $\delta$ tends to decrease for heavier $m_0$.
It is because it would be more difficult to amplify heavier particles.
This feature is also verified in \cite{Pappas}.
In the case of 4-dimensional scattering, the presence of the mass term makes it possible to form a bound state of the black hole and
the scattered particle, only when $\omega<m_0$.
In this case the instability of the black hole gets stronger as $m_0$ increases because the depth of the
effective potential well becomes larger, as pointed out in \cite{Yoshida}.
But in the current work we only consider the region of $\omega>m_0$ to make use of the expansion like Eq.\ (\ref{tE}),
with the assumption of the mirror boundary condition.
Thus we cannot expect a smiliar behavior to that of \cite{Yoshida}.
\par
For all the cases considered in this paper, the largest value of $\delta_{\rm max}$ is $0.606$ GeV 
(brane emission of $(jm)=(11)$ for $a_*=1.5$ and $m_0=120$ GeV), 
which is smaller than the largest $\delta_{\rm max}$ of \cite{jplee}, $0.772$ GeV
(brane emission of $(jm)=(11)$ for $a_*=1.5$ and $m_0=m_\pi$ GeV).
And in both cases of bulk emission and brane emission, 
one can easily distinguish the modes $(j\ell m)=(101)$ from $(202)$, or $(jm)=(11)$ from $(22)$,
since for $(101)$ or $(11)$ modes $\omega_c r_h<0.5$ while for $(202)$ or $(22)$ modes $\omega_c r_h>0.5$.
Actually  $\omega_c r_h<0.5$ for $(101)$ or $(11)$ modes is obvious because 
\begin{equation}
 \omega_{\rm max}r_h=\frac{a_*m}{1+a_*^2}\le\frac{1}{2}m~,
\end{equation}
from Eq.\ (\ref{wmax}).
\par
The $e$-folding time $t_c=1/\delta_{\rm max}$ is a quantity such that for every $t_c$ for $\omega=\omega_c$
the field amplitude is amplified by $e$-times and the energy gets larger by an amount of $e^2=7.4$. 
Kerr instability is comparable to Hawking radiation when $e$-folding time is shorter than the Hawking evaporation lifetime \cite{Dolan}.
For the bulk emission $t_c\gtrsim 10^{-22}\sec$ while $t_c\gtrsim 10^{-24}\sec$ for the brane emission.
These values are quite large compared with the typical mini black hole lifetime \cite{Kanti, Argyres}
$\tau_{\rm BH}\sim (1.7-0.5)\times 10^{-26}\sec$.
For this value of the black hole lifetime, the amplification by the superradiance is very small, $\lesssim\exp(1/100)$ at best.
But it is not impossible that $\tau_{\rm BH}$ could be much longer, say, $\tau_{\rm BH}\sim 10^{-17}\sec$ \cite{Casadio}.
Even for the bulk emission case the scattered particles can experience large amplifications if the black hole lifetime is long enough.
Thus there is a chance to observe the black hole bombs at the LHC near future.
\par
As in the brane emission case we can estimated the relative energy extracted from the black hole 
when the mirror is put near its minimum value \cite{jplee}.
From the first law of thermodynamics of black holes, one has
\begin{equation}
 \frac{\Delta\MBH}{\MBH}=\left(\frac{2}{n+2}\right)\frac{a_*}{1+a_*^2}\Delta a_*~.
\label{DeltaM}
\end{equation}
As the scattering process proceeds, the black hole loses its rotational energy and $a_*$ gets smaller.
Note that $\r0min$ is minimum at $a_*=1$ (see Eqs.\ (\ref{wmax}) and(\ref{r0min})) and becomes larger for smaller $a_*$.
For example, consider the bulk emission of mode $(j\ell m)=(202)$ with $m_0=m_\pi$ and $a_*=0.5$.
See the Table \ref{bulk202pi}.
In this case $\r0min=8.73 r_h$, so put the mirror at $r_0=9.00 r_h$ at the beginning.
During the scattering $a_*$ decreases while $\r0min$ increases, and eventually there is a point where
$\r0min(a_*)=9.00 r_h$.
If $\r0min$ exceeds the mirror location, $r_0=9.00 r_h$, the superradiance stops.
This happens when $a_*=0.475$ where $\r0min(a_*=0.475)=9.00 r_h$.
For this process the energy fraction transferred from the black hole is $\Delta\MBH/\MBH\simeq 0.5\%$ for $n=2$ from
Eq.\ (\ref{DeltaM}).
This is a half value of the fraction for the brane emission case considered in \cite{jplee}.
Similar calculation shows that for the case of $(j\ell m)=(101)$ with $m_0=m_\pi$, $a_*=0.5$ in the bulk emission,
$\Delta\MBH/\MBH\simeq 0.6\%$.
If one puts the mirror far enough from $\r0min$, one can extract much more energy from the black hole \cite{Cardoso}.

\section{conclusions}
In this paper the superradiant scattering by rotating black holes in higher dimensions with mirror boundary conditions 
(the so called {\em black-hole bomb}) is investigated.
The work is an extension of the previous one \cite{jplee} in that in this version the bulk emission case is also analyzed, 
and other modes of $(j\ell m)$ or $(jm)$ are compared, and different mass effects are considered.
Among various cases in the analyses we found that the largest value of $\delta_{\rm max}$ is obtained for the 
brane emission with $(jm)=(11)$ for smaller scalar mass and larger angular momentum of the black hole.
In the bulk emission $\delta_{\rm max}$ is suppressed by a factor of $\sim 10^{-2}$ or more.
For the brane emission the shortest $e$-folding time is about $\sim 10^{-24} \sec$ which is much longer than the typical 
lifetime of mini black holes.
But it is possible that the long-lived mini black holes might exist and could provide a chance for forming the black-hole bombs.
Any technical details on how to realize the mirror to bounce off the scattered particles back to the black hole are beyond the 
scope of this work and are not considered here.

\begin{acknowledgments}
This work was supported by the Basic Science Research Program through the National Research Foundation of Korea (NRF) 
funded by the Korean Ministry of Education, Science and Technology (2009-00888396).
\end{acknowledgments}

\begin{figure}
\begin{tabular}{cc}
\includegraphics{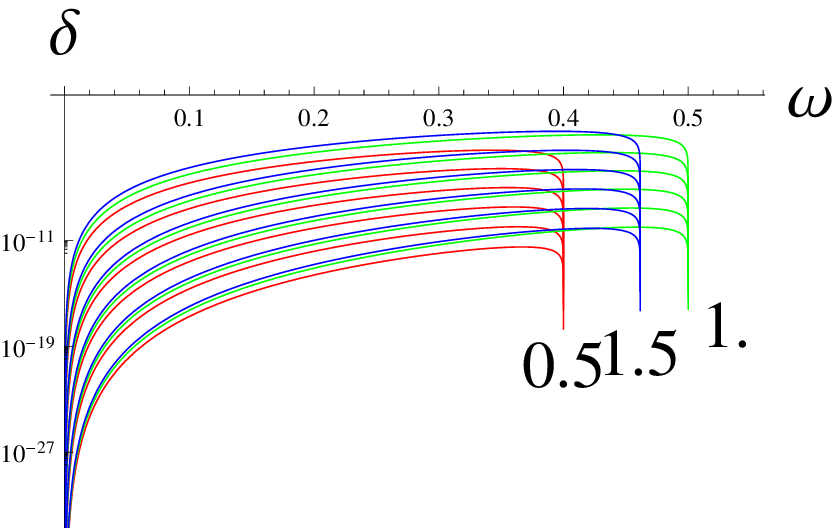}&\includegraphics{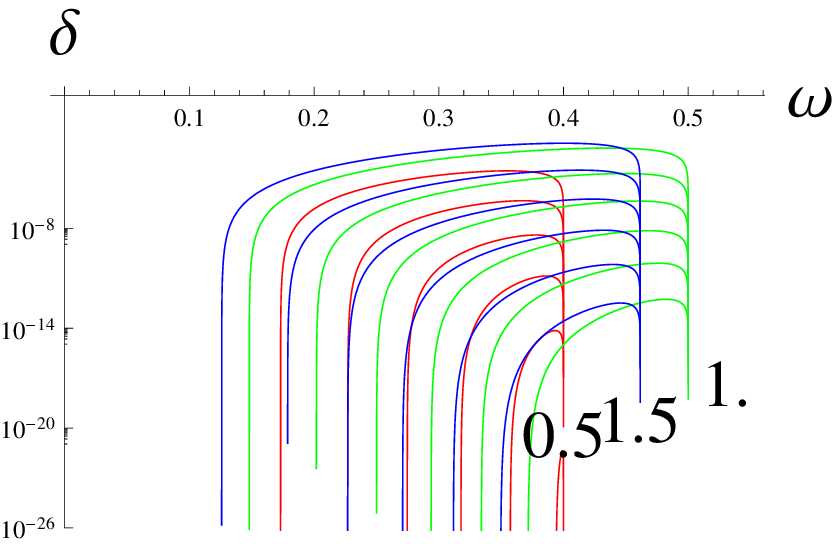}\\
(a) $(j\ell m)=(101)$, $m_\phi=m_\pi$. & (b) $(j\ell m)=(101)$, $m_\phi=120$ GeV.\\
\includegraphics{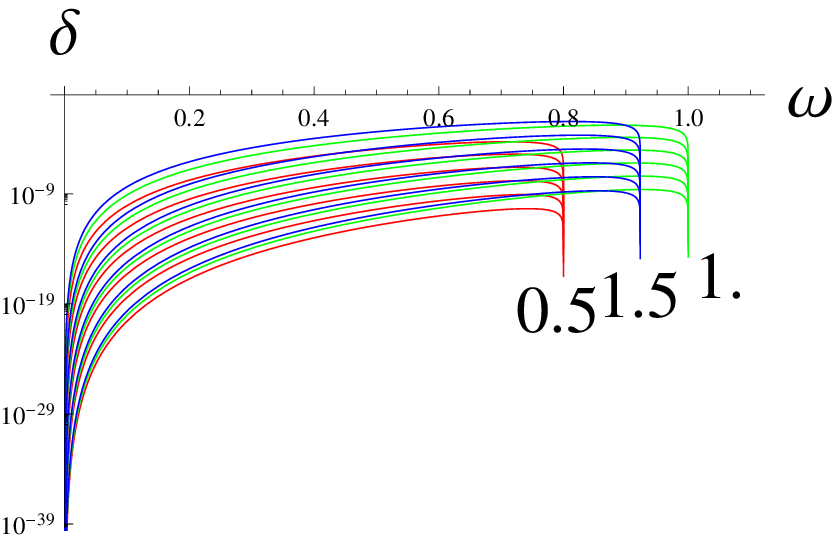}&\includegraphics{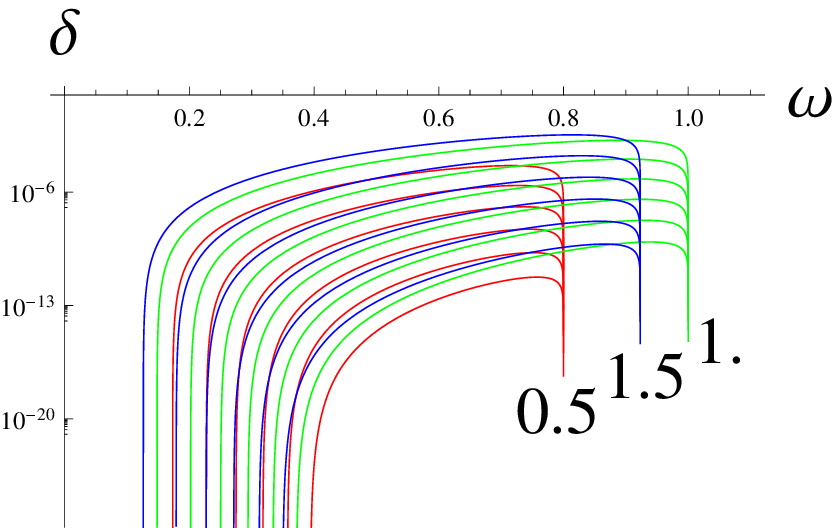}\\
(c) $(j\ell m)=(202)$, $m_\phi=m_\pi$. & (d) $(j\ell m)=(202)$, $m_\phi=120$ GeV.\\
\end{tabular}
\caption{\label{bulk}Plots of $\delta$ in GeV as a function of $\omega$ in units of $1/r_h$ 
for the bulk emission for $m_0=m_\pi$, $m_0=120$ GeV for the modes $(j\ell m)=(101), (202)$.
Numbers are shown to denote the value of $a_*$.
Each graphs in the stacks for different $a_*$ corresponds to $n=2, \cdots, 7$ from top to bottom.}
\end{figure}
\begin{figure}
\begin{tabular}{cc}
\includegraphics{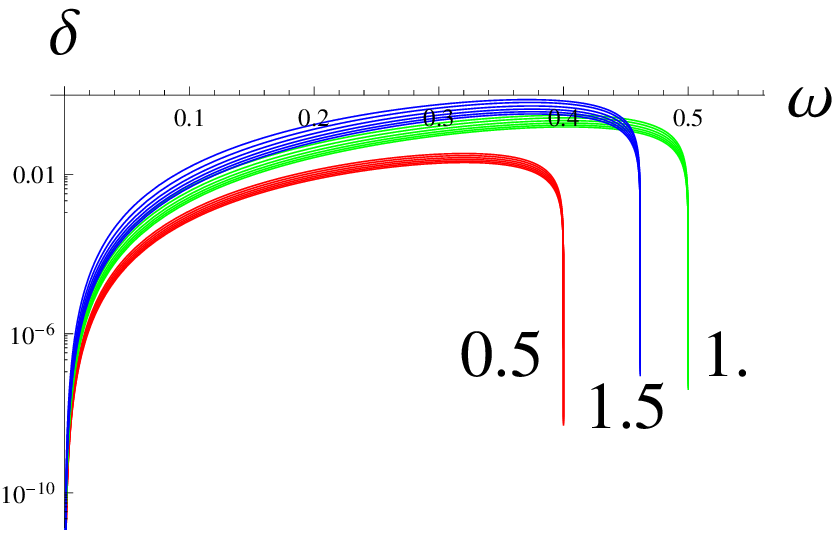}&\includegraphics{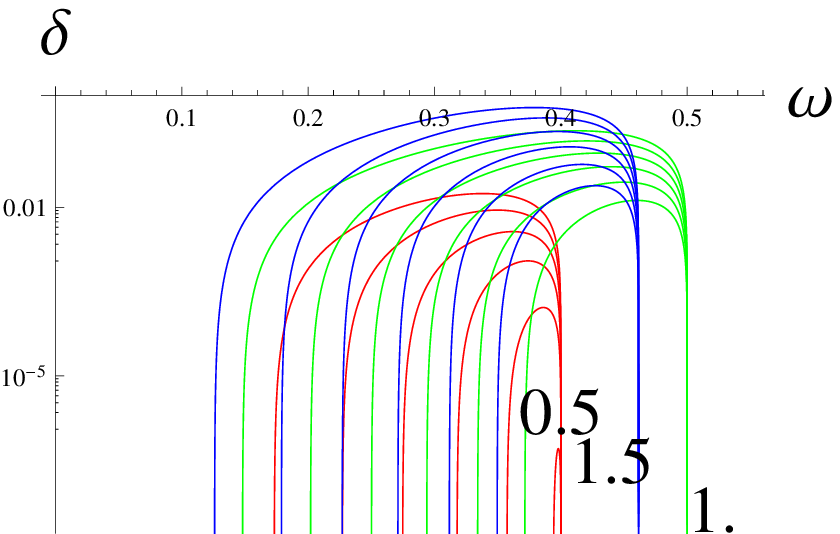}\\
(a) $(jm)=(11)$, $m_\phi=m_\pi$. & (b) $(jm)=(11)$, $m_\phi=120$ GeV.\\
\includegraphics{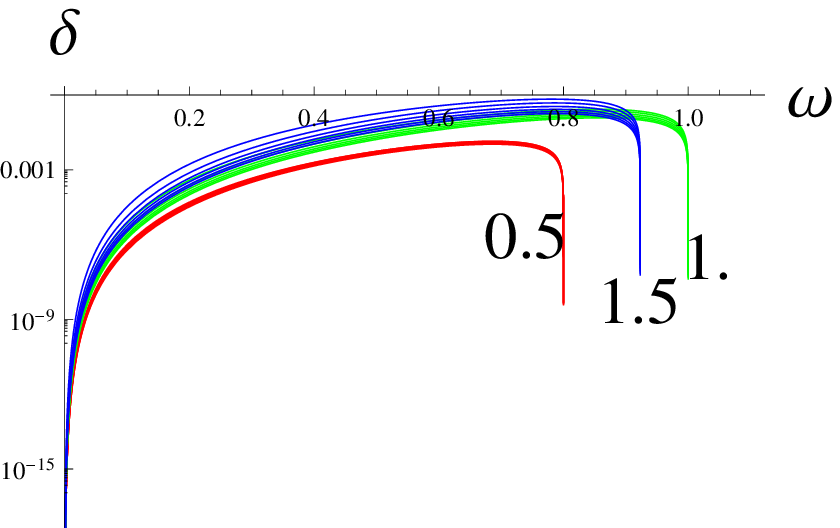}&\includegraphics{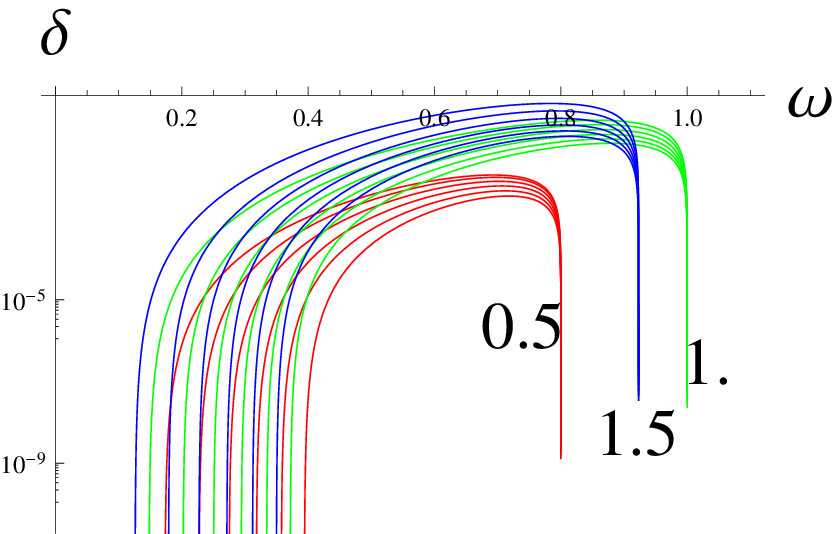}\\
(c) $(jm)=(22)$, $m_\phi=m_\pi$. & (d) $(j m)=(22)$, $m_\phi=120$ GeV.\\
\end{tabular}
\caption{\label{brane}Plots of $\delta$ as a function of $\omega$ in units of $1/r_h$ 
for the brane emission for $m_0=m_\pi$, $m_0=120$ GeV for the modes $(j\ell m)=(101), (202)$.
Numbers are shown to denote the value of $a_*$.
Each stacks of graphs corresponds to $n=2, \cdots, 7$ from top to bottom, with one exception. See the text.}
\end{figure}

\begin{table}
\begin{tabular}{cc|cccccc}\hline
&$n$ &$2$&$3$&$4$&$5$&$6$&$7$\\\hline
&$0.5~$&$~2.85~$&$~3.73~$&$~4.52~$&$~5.23~$&$~5.87~$&$~6.48~$\\
$r_h~$&$1.0~$&$2.43$&$3.32$&$4.11$&$4.83$&$5.49$&$6.11$\\
($10^{-4}$ fm)&$1.5~$&$2.07$&$2.94$&$3.73$&$4.46$&$5.13$&$5.75$\\\hline
&$0.5~$&$~6.72\times 10^{-5}~$&$~2.71\times 10^{-6}~$&$~1.00\times 10^{-7}~$&$~3.43\times 10^{-9}~$&$~1.10\times 10^{-10}~$&$~3.27\times 10^{-12}~$\\
$\delta_{\rm max}$&$1.0~$&$9.74\times 10^{-4}$&$4.45\times 10^{-5}$&$1.88\times 10^{-6}$&$7.55\times 10^{-8}$&$2.86\times 10^{-9}$&$1.02\times 10^{-10}$\\
(GeV)&$1.5~$&$1.80\times 10^{-3}$&$6.59\times 10^{-5}$&$2.31\times 10^{-6}$&$7.94\times 10^{-8}$&$2.63\times 10^{-9}$&$8.35\times 10^{-11}$\\\hline
&$0.5~$&$9.79\times 10^{-21}$&$2.43\times 10^{-19}$&$6.58\times 10^{-18}$&$1.92\times 10^{-16}$&$6.01\times 10^{-15}$&$2.12\times 10^{-13}$\\
$t_c$ &$1.0~$&$6.76\times 10^{-22}$&$1.48\times 10^{-20}$&$3.50\times 10^{-19}$&$8.72\times 10^{-18}$&$2.30\times 10^{-16}$&$6.43\times 10^{-15}$\\
(sec) &$1.5~$&$~3.66\times 10^{-22}~$&$~9.98\times 10^{-21}~$&$~2.85\times 10^{-19}~$&$~8.29\times 10^{-18}~$&$~2.50\times 10^{-16}~$&$~7.88\times 10^{-15}~$\\\hline
&$0.5~$&$0.343$&$0.350$&$0.356$&$0.360$&$0.364$&$0.367$\\
$\omega_c r_h$ &$1.0~$&$0.428$&$0.437$&$0.444$&$0.450$&$0.454$&$0.458$\\
 &$1.5~$&$~0.394~$&$~0.403~$&$~0.410~$&$~0.415~$&$~0.419~$&$~0.423~$\\\hline
&$0.5~$&$16.8$&$18.2$&$19.6$&$21.1$&$22.5$&$23.9$\\
$r_0(\omega_c)/r_h$ &$1.0~$&$13.5$&$14.6$&$15.8$&$16.9$&$18.0$&$19.2$\\
 &$1.5~$&$~14.8~$&$~15.9~$&$~17.1~$&$~18.3~$&$~19.6~$&$~20.8~$\\\hline
&$0.5~$&$14.4$&$16.0$&$17.5$&$19.0$&$20.5$&$21.9$\\
$\r0min/r_h$ &$1.0~$&$11.5$&$12.8$&$14.0$&$15.2$&$16.4$&$17.5$\\
 &$1.5~$&$~12.5~$&$~13.8~$&$~15.1~$&$~16.4~$&$~17.7~$&$~19.0~$\\\hline
\end{tabular}
\caption{\label{bulk101pi}Some parameters for different values of $n=2,3,\cdots,7$ and $a_*=0.5,~1.0,~1.5$ 
for $(j\ell m)=(101)$, $m_0=m_\pi$ (bulk emission).}
\end{table}

\begin{table}
\begin{tabular}{cc|cccccc}\hline
&$n$ &$2$&$3$&$4$&$5$&$6$&$7$\\\hline
&$0.5~$&$~2.88\times 10^{-5}~$&$~4.60\times 10^{-7}~$&$~4.00\times 10^{-9}~$&$~1.38\times 10^{-11}~$&$~7.24\times 10^{-15}~$&$~4.23\times 10^{-22}~$\\
$\delta_{\rm max}$&$1.0~$&$6.69\times 10^{-4}$&$1.98\times 10^{-5}$&$4.41\times 10^{-7}$&$7.25\times 10^{-9}$&$8.13\times 10^{-11}$&$5.48\times 10^{-13}$\\
(GeV)&$1.5~$&$1.31\times 10^{-3}$&$3.16\times 10^{-5}$&$5.78\times 10^{-7}$&$7.74\times 10^{-9}$&$6.80\times 10^{-11}$&$3.27\times 10^{-13}$\\\hline
&$0.5~$&$2.29\times 10^{-20}$&$1.43\times 10^{-18}$&$1.64\times 10^{-16}$&$4.78\times 10^{-14}$&$9.09\times 10^{-11}$&$1.56\times 10^{-3}$\\
$t_c$ &$1.0~$&$9.83\times 10^{-22}$&$3.33\times 10^{-20}$&$1.49\times10^{-18}$&$9.08\times 10^{-17}$&$8.10\times 10^{-15}$&$1.20\times 10^{-12}$\\
(sec) &$1.5~$&$~5.01\times 10^{-22}~$&$~2.08\times 10^{-20}~$&$~1.14\times 10^{-18}~$&$~8.50\times 10^{-17}~$&$~9.68\times 10^{-15}~$&$~2.01\times 10^{-12}~$\\\hline
&$0.5~$&$0.355$&$0.368$&$0.378$&$0.386$&$0.393$&$0.399$\\
$\omega_c r_h$ &$1.0~$&$0.435$&$0.449$&$0.459$&$0.468$&$0.476$&$0.482$\\
 &$1.5~$&$~0.400~$&$~0.413~$&$~0.423~$&$~0.432~$&$~0.440~$&$~0.446~$\\\hline
&$0.5~$&$18.6$&$22.0$&$26.9$&$34.6$&$50.0$&$138$\\
${r_0(\omega_c)}/{r_h}$ &$1.0~$&$14.1$&$15.9$&$18.2$&$20.9$&$24.1$&$28.6$\\
 &$1.5~$&$~15.3~$&$~17.2~$&$~19.6~$&$~22.6~$&$~26.4~$&$~31.7~$\\\hline
&$0.5~$&$16.0$&$19.4$&$24.0$&$31.2$&$45.5$&$125.9$\\
$\r0min/r_h$ &$1.0~$&$12.1$&$14.0$&$16.1$&$18.8$&$22.0$&$26.2$\\
 &$1.5~$&$~13.0~$&$~15.0~$&$~17.4~$&$~20.3~$&$~24.0~$&$~29.1~$\\\hline
\end{tabular}
\caption{\label{bulk101120}Some parameters for different values of $n=2,3,\cdots,7$ and $a_*=0.5,~1.0,~1.5$ 
for $(j\ell m)=(101)$, $m_0=120$ GeV (bulk emission).}
\end{table}

\begin{table}
\begin{tabular}{cc|cccccc}\hline
&$n$ &$2$&$3$&$4$&$5$&$6$&$7$\\\hline
&$0.5~$&$~5.55\times 10^{-5}~$&$~4.12\times 10^{-6}~$&$~2.66\times 10^{-7}~$&$~1.58\times 10^{-8}~$&$~8.76\times 10^{-10}~$&$~4.55\times 10^{-11}~$\\
$\delta_{\rm max}$&$1.0~$&$1.78\times 10^{-3}$&$1.37\times 10^{-4}$&$9.63\times 10^{-6}$&$6.44\times 10^{-7}$&$4.12\times 10^{-8}$&$2.52\times 10^{-9}$\\
(GeV)&$1.5~$&$2.12\times 10^{-3}$&$9.40\times 10^{-5}$&$4.08\times 10^{-6}$&$1.77\times 10^{-7}$&$7.51\times 10^{-9}$&$3.10\times 10^{-10}$\\\hline
&$0.5~$&$1.19\times 10^{-20}$&$1.60\times 10^{-19}$&$2.47\times 10^{-18}$&$4.16\times 10^{-17}$&$7.52\times 10^{-16}$&$1.45\times 10^{-14}$\\
$t_c$ &$1.0~$&$3.71\times 10^{-22}$&$4.80\times 10^{-21}$&$6.83\times 10^{-20}$&$1.02\times 10^{-18}$&$1.60\times 10^{-17}$&$2.61\times 10^{-16}$\\
(sec) &$1.5~$&$~3.10\times 10^{-22}~$&$~7.00\times 10^{-21}~$&$~1.61\times 10^{-19}~$&$~3.73\times 10^{-18}~$&$~8.77\times 10^{-17}~$&$~2.13\times 10^{-15}~$\\\hline
&$0.5~$&$0.711$&$0.720$&$0.727$&$0.733$&$0.738$&$0.743$\\
$\omega_c r_h$ &$1.0~$&$0.885$&$0.897$&$0.907$&$0.915$&$0.922$&$0.928$\\
 &$1.5~$&$~0.667~$&$~0.667~$&$~0.667~$&$~0.667~$&$~0.667~$&$~0.667~$\\\hline
&$0.5~$&$9.85$&$10.6$&$11.3$&$12.0$&$12.7$&$13.4$\\
${r_0(\omega_c)}/{r_h}$ &$1.0~$&$8.00$&$8.54$&$9.09$&$9.64$&$10.2$&$10.7$\\
 &$1.5~$&$~10.7~$&$~11.5~$&$~12.4~$&$~13.2~$&$~14.1~$&$~15.0~$\\\hline
&$0.5~$&$8.73$&$9.49$&$10.2$&$11.0$&$11.7$&$12.4$\\
$\r0min/r_h$ &$1.0~$&$6.99$&$7.59$&$8.18$&$8.77$&$9.36$&$9.94$\\
 &$1.5~$&$~10.5~$&$~11.4~$&$~12.3~$&$~13.2~$&$~14.0~$&$14.9$\\\hline
\end{tabular}
\caption{\label{bulk202pi}Some parameters for different values of $n=2,3,\cdots,7$ and $a_*=0.5,~1.0,~1.5$ 
for $(j\ell m)=(202)$, $m_0=m_\pi$ (bulk emission).}
\end{table}

\begin{table}
\begin{tabular}{cc|cccccc}\hline
&$n$ &$2$&$3$&$4$&$5$&$6$&$7$\\\hline
&$0.5~$&$~4.35\times 10^{-5}~$&$~2.57\times 10^{-6}~$&$~1.25\times 10^{-7}~$&$~5.15\times 10^{-9}~$&$~1.83\times 10^{-10}~$&$~5.62\times 10^{-12}~$\\
$\delta_{\rm max}$&$1.0~$&$1.59\times 10^{-3}$&$1.09\times 10^{-4}$&$6.56\times 10^{-6}$&$3.60\times 10^{-7}$&$1.81\times 10^{-8}$&$8.34\times 10^{-10}$\\
(GeV)&$1.5~$&$2.01\times 10^{-3}$&$8.47\times 10^{-5}$&$3.46\times 10^{-6}$&$1.39\times 10^{-7}$&$5.46\times 10^{-9}$&$2.05\times 10^{-10}$\\\hline
&$0.5~$&$1.51\times 10^{-20}$&$2.56\times 10^{-19}$&$5.29\times 10^{-18}$&$1.28\times 10^{-16}$&$3.59\times 10^{-15}$&$1.17\times 10^{-13}$\\
$t_c$ &$1.0~$&$4.15\times 10^{-22}$&$6.03\times 10^{-21}$&$1.00\times 10^{-19}$&$1.83\times 10^{-18}$&$3.63\times 10^{-17}$&$7.90\times 10^{-16}$\\
(sec) &$1.5~$&$~3.28\times 10^{-22}~$&$~7.77\times 10^{-21}~$&$~1.90\times 10^{-19}~$&$~4.72\times 10^{-18}~$&$~1.21\times 10^{-16}~$&$~3.21\times 10^{-15}~$\\\hline
&$0.5~$&$0.716$&$0.727$&$0.737$&$0.745$&$0.751$&$0.757$\\
$\omega_c r_h$ &$1.0~$&$0.888$&$0.902$&$0.914$&$0.923$&$0.931$&$0.938$\\
 &$1.5~$&$~0.678~$&$~0.690~$&$~0.704~$&$~0.720~$&$~0.736~$&$~0.753~$\\\hline
&$0.5~$&$10.1$&$11.0$&$12.0$&$13.0$&$14.2$&$15.4$\\
${r_0(\omega_c)}/{r_h}$ &$1.0~$&$8.08$&$8.71$&$9.37$&$10.1$&$10.8$&$11.6$\\
 &$1.5~$&$~10.7~$&$~11.5~$&$~12.4~$&$~13.2~$&$~14.1~$&$~15.0~$\\\hline
&$0.5~$&$8.95$&$9.89$&$10.9$&$11.9$&$13.1$&$14.3$\\
$\r0min/r_h$ &$1.0~$&$7.07$&$7.75$&$8.45$&$9.18$&$9.93$&$10.7$\\
 &$1.5~$&$~10.5~$&$~11.4~$&$~12.3~$&$~13.2~$&$~14.0~$&$~14.9~$\\\hline
\end{tabular}
\caption{\label{bulk202120}Some parameters for different values of $n=2,3,\cdots,7$ and $a_*=0.5,~1.0,~1.5$ 
for $(j\ell m)=(202)$, $m_0=120$ GeV (bulk emission).}
\end{table}

\begin{table}
\begin{tabular}{cc|cccccc}\hline
&$n$ &$2$&$3$&$4$&$5$&$6$&$7$\\\hline
&$0.5~$&$~1.78\times 10^{-2}~$&$~9.02\times 10^{-3}~$&$~3.73\times 10^{-3}~$&$~1.12\times 10^{-3}~$&$~1.66\times 10^{-4}~$&$~5.05\times 10^{-7}~$\\
$\delta_{\rm max}$&$1.0~$&$2.34\times 10^{-1}$&$1.54\times 10^{-1}$&$9.35\times 10^{-2}$&$5.35\times 10^{-2}$&$2.84\times 10^{-2}$&$1.34\times 10^{-2}$\\
(GeV)&$1.5~$&$6.06\times 10^{-1}$&$3.96\times 10^{-1}$&$2.27\times 10^{-1}$&$1.21\times 10^{-1}$&$5.88\times 10^{-2}$&$2.46\times 10^{-2}$\\\hline
&$0.5~$&$3.71\times 10^{-23}$&$7.30\times 10^{-23}$&$1.77\times 10^{-22}$&$5.87\times 10^{-22}$&$3.97\times 10^{-21}$&$1.30\times 10^{-18}$\\
$t_c$ &$1.0~$&$2.81\times 10^{-24}$&$4.27\times 10^{-24}$&$7.04\times10^{-24}$&$1.23\times 10^{-23}$&$2.32\times 10^{-23}$&$4.91\times 10^{-23}$\\
(sec) &$1.5~$&$~1.09\times 10^{-24}~$&$~1.66\times 10^{-24}~$&$~2.90\times 10^{-24}~$&$~5.44\times 10^{-24}~$&$~1.12\times 10^{-23}~$&$~2.68\times 10^{-23}~$\\\hline
&$0.5~$&$0.338$&$0.350$&$0.362$&$0.374$&$0.386$&$0.398$\\
$\omega_c r_h$ &$1.0~$&$0.411$&$0.420$&$0.430$&$0.440$&$0.450$&$0.460$\\
 &$1.5~$&$~0.379~$&$~0.388~$&$~0.397~$&$~0.407~$&$~0.417~$&$~0.427~$\\\hline
&$0.5~$&$15.5$&$16.9$&$19.1$&$22.8$&$30.8$&$78.8$\\
${r_0(\omega_c)}/{r_h}$ &$1.0~$&$11.9$&$12.3$&$13.0$&$13.8$&$15.0$&$16.7$\\
 &$1.5~$&$~12.8~$&$~13.3~$&$~14.0~$&$~15.0~$&$~16.4~$&$~18.5~$\\\hline
&$0.5~$&$12.5$&$13.6$&$15.5$&$18.5$&$25.0$&$64.5$\\
$\r0min/r_h$ &$1.0~$&$9.41$&$9.82$&$10.4$&$11.1$&$12.1$&$13.4$\\
 &$1.5~$&$~10.1~$&$~10.6~$&$~11.2~$&$~12.0~$&$~13.2~$&$~14.9~$\\\hline
\end{tabular}
\caption{\label{brane11120}Some parameters for different values of $n=2,3,\cdots,7$ and $a_*=0.5,~1.0,~1.5$ 
for $(jm)=(11)$, $m_0=120$ GeV (brane emission).}
\end{table}
\begin{table}
\begin{tabular}{cc|cccccc}\hline
&$n$ &$2$&$3$&$4$&$5$&$6$&$7$\\\hline
&$0.5~$&$~1.37\times 10^{-2}~$&$~1.39\times 10^{-2}~$&$~1.32\times 10^{-2}~$&$~1.23\times 10^{-2}~$&$~1.14\times 10^{-2}~$&$~1.06\times 10^{-2}~$\\
$\delta_{\rm max}$&$1.0~$&$2.69\times 10^{-1}$&$2.27\times 10^{-1}$&$1.87\times 10^{-1}$&$1.58\times 10^{-1}$&$1.37\times 10^{-1}$&$1.21\times 10^{-1}$\\
(GeV)&$1.5~$&$5.19\times 10^{-1}$&$3.48\times 10^{-1}$&$2.46\times 10^{-1}$&$1.87\times 10^{-1}$&$1.50\times 10^{-1}$&$1.25\times 10^{-1}$\\\hline
&$0.5~$&$4.80\times 10^{-23}$&$4.73\times 10^{-23}$&$5.00\times 10^{-23}$&$5.37\times 10^{-23}$&$5.77\times 10^{-23}$&$6.18\times 10^{-23}$\\
$t_c$ &$1.0~$&$2.45\times 10^{-24}$&$2.90\times 10^{-24}$&$3.52\times10^{-24}$&$4.17\times 10^{-24}$&$4.82\times 10^{-24}$&$5.45\times 10^{-24}$\\
(sec) &$1.5~$&$~1.27\times 10^{-24}~$&$~1.89\times 10^{-24}~$&$~2.68\times 10^{-24}~$&$~3.53\times 10^{-24}~$&$~4.40\times 10^{-24}~$&$~5.27\times 10^{-24}~$\\\hline
&$0.5~$&$0.686$&$0.686$&$0.686$&$0.686$&$0.686$&$0.686$\\
$\omega_c r_h$ &$1.0~$&$0.852$&$0.854$&$0.855$&$0.855$&$0.855$&$0.855$\\
 &$1.5~$&$~0.667~$&$~0.667~$&$~0.667~$&$~0.667~$&$~0.667~$&$~0.667~$\\\hline
&$0.5~$&$8.44$&$8.44$&$8.44$&$8.44$&$8.44$&$8.44$\\
${r_0(\omega_c)}/{r_h}$ &$1.0~$&$6.94$&$6.93$&$6.92$&$6.92$&$6.92$&$6.92$\\
 &$1.5~$&$~8.95~$&$~8.95~$&$~8.95~$&$~8.95~$&$~8.95~$&$~8.95~$\\\hline
&$0.5~$&$7.20$&$7.20$&$7.20$&$7.20$&$7.20$&$7.20$\\
$\r0min/r_h$ &$1.0~$&$5.76$&$5.76$&$5.76$&$5.76$&$5.76$&$5.76$\\
 &$1.5~$&$~8.65~$&$~8.65~$&$~8.65~$&$~8.65~$&$~8.65~$&$~8.65~$\\\hline
\end{tabular}
\caption{\label{brane22pi}Some parameters for different values of $n=2,3,\cdots,7$ and $a_*=0.5,~1.0,~1.5$ 
for $(jm)=(22)$, $m_0=m_\pi$ (brane emission).}
\end{table}

\begin{table}
\begin{tabular}{cc|cccccc}\hline
&$n$ &$2$&$3$&$4$&$5$&$6$&$7$\\\hline
&$0.5~$&$~1.12\times 10^{-2}~$&$~9.84\times 10^{-3}~$&$~7.87\times 10^{-3}~$&$~6.08\times 10^{-3}~$&$~4.60\times 10^{-3}~$&$~3.44\times 10^{-3}~$\\
$\delta_{\rm max}$&$1.0~$&$2.46\times 10^{-1}$&$1.92\times 10^{-1}$&$1.45\times 10^{-1}$&$1.10\times 10^{-1}$&$8.55\times 10^{-2}$&$6.71\times 10^{-2}$\\
(GeV)&$1.5~$&$4.93\times 10^{-1}$&$3.18\times 10^{-1}$&$2.14\times 10^{-1}$&$1.52\times 10^{-1}$&$1.14\times 10^{-1}$&$8.75\times 10^{-2}$\\\hline
&$0.5~$&$5.85\times 10^{-23}$&$6.69\times 10^{-23}$&$8.31\times 10^{-23}$&$1.08\times 10^{-22}$&$1.43\times 10^{-22}$&$1.92\times 10^{-22}$\\
$t_c$ &$1.0~$&$2.68\times 10^{-24}$&$3.42\times 10^{-24}$&$4.55\times10^{-24}$&$5.96\times 10^{-24}$&$7.70\times 10^{-24}$&$9.81\times 10^{-24}$\\
(sec) &$1.5~$&$~1.34\times 10^{-24}~$&$~2.07\times 10^{-24}~$&$~3.08\times 10^{-24}~$&$~4.32\times 10^{-24}~$&$~5.78\times 10^{-24}~$&$~7.52\times 10^{-24}~$\\\hline
&$0.5~$&$0.692$&$0.697$&$0.702$&$0.707$&$0.712$&$0.717$\\
$\omega_c r_h$ &$1.0~$&$0.856$&$0.861$&$0.865$&$0.869$&$0.873$&$0.878$\\
 &$1.5~$&$~0.678~$&$~0.690~$&$~0.704~$&$~0.720~$&$~0.736~$&$~0.753~$\\\hline
&$0.5~$&$8.64$&$8.78$&$8.96$&$9.16$&$9.39$&$9.65$\\
${r_0(\omega_c)}/{r_h}$ &$1.0~$&$7.01$&$7.06$&$7.13$&$7.22$&$7.31$&$7.41$\\
 &$1.5~$&$~8.95~$&$~8.95~$&$~8.95~$&$~8.95~$&$~8.95~$&$~8.95~$\\\hline
&$0.5~$&$7.38$&$7.51$&$7.67$&$7.85$&$8.05$&$8.28$\\
$\r0min/r_h$ &$1.0~$&$5.83$&$5.88$&$5.95$&$6.03$&$6.11$&$6.21$\\
 &$1.5~$&$~8.65~$&$~8.65~$&$~8.65~$&$~8.65~$&$~8.65~$&$~8.65~$\\\hline
\end{tabular}
\caption{\label{brane22120}Some parameters for different values of $n=2,3,\cdots,7$ and $a_*=0.5,~1.0,~1.5$ 
for $(jm)=(22)$, $m_0=120$ GeV (brane emission).}
\end{table}

\end{document}